# Defect-engineered graphene for bulk supercapacitors with high-energy and power densities

Jingyi Zhu[1], Anthony Childress[1], Mehmet Karakaya[1], Anurag Srivastava[2], Ye Lin[3], Apparao M. Rao[1,*], and Ramakrishna Podila[1,*]

[1] Department of Physics and Astronomy, Clemson Nanomaterials Center and COMSET, Clemson University, Clemson, SC 29634, USA

[2] ABV-Indian Institute of Information Technology and Management, Gwalior (M.P.) 474010, INDIA

[3] Department of Mechanical Engineering, University of South Carolina, Columbia, SC 29208, USA

*Corresponding author: rpodila@g.clemson.edu, arao@g.clemson.edu

**The development of high-energy and high-power density supercapacitors (SCs) is critical for enabling next-generation energy storage applications. Nanocarbons are excellent SC electrode materials due to their economic viability, high-surface area, and high stability. Although nanocarbons have high theoretical surface area and hence high double layer capacitance, the net amount of energy stored in nanocarbon-SCs is much below theoretical limits due to two inherent bottlenecks: i) their low quantum capacitance and ii) limited ion-accessible surface area. Here, we demonstrate that defects in graphene could be effectively used to mitigate these bottlenecks by drastically increasing the quantum capacitance and opening new channels to facilitate ion diffusion in otherwise closed interlayer spaces. Our results support the emergence of a new energy paradigm in SCs with 250% enhancement in double layer capacitance beyond the theoretical limit. Furthermore, we demonstrate prototype defect engineered bulk SC devices with energy densities 500% higher than state-of-the-art commercial SCs without compromising the power density.**

## Introduction

Supercapacitors (SCs) are novel electrochemical devices that store energy through reversible adsorption of ionic species from an electrolyte on highly porous electrode surfaces. SCs are highly durable (lifetime >10,000 cycles) with power densities (10 kW/kg) that are an order of magnitude larger than batteries. But the low energy density (10 Wh/kg) of SCs[1] relative to batteries precludes their use in practical applications despite their ability to withstand >10,000 cycles. Graphene-based nanocarbons are ideal electrode materials for SCs due to their low cost, high stability, and high specific surface area. Indeed, an outstanding characteristic of single-layer graphene is its high specific surface area ~2675 m²/g, which sets an upper limit for electrical double layer capacitance ($C_{dl}$) ~21 µF/cm² (~550 F/g).[1–4] Notwithstanding this theoretical limit, there are two intrinsic bottlenecks that are impeding the emergence of high energy density SC devices: i) typically only 50-70% of the theoretical surface area is accessible to ionic species from the electrolyte, which limits the overall capacitance (10-15 µF/cm²) and leads to low energy density, and ii) although the total energy that can be harnessed from a SC device depends predominantly on ion-accessible surface area, it is not the only factor. The presence of the so-called small quantum capacitance ($C_Q$) in series for nanocarbon electrodes, arising from their low electronic density of states at the Fermi level (DOS($E_F$)), overwhelms the high $C_{dl}$ further reducing the already limited capacitance and low energy density.[5–7]

While the efforts to increase energy density have been focused either on increasing the active surface area or the addition of pseudo-capacitance through redox active materials, there is a clear lack of methodologies to simultaneously address the inherent challenges described above. Here, we experimentally show that engineered defects in graphene can alleviate these bottlenecks resulting in a new paradigm of energy storage beyond the predicted theoretical limits. Defects are often perceived as performance limiters in graphene. Yet, our experimental findings conclusively demonstrate that controllably induced defects in specific configurations can achieve 250% enhancement (~50 µF/cm²) in measurable capacitance of few-layer graphene (FLG). Our detailed density function (DFT) theory calculations show that the N-dopants in pyrrolic configuration result in a high DOS($E_F$) and thereby mitigate the influence of $C_Q$. Furthermore, the inter-layer spaces in FLG can be accessed by tetraethylammonium (TEA⁺) ions effectively through defect-induced pores leading to increased charge storage. More importantly, we show that these high-capacitances can be extended to coin-cell devices based on FLG foams that result in energy densities at least five times higher than the conventional activated carbon SCs.

# Results

**Figure 1: The interaction of electrolyte ions with defect-induced pores.**

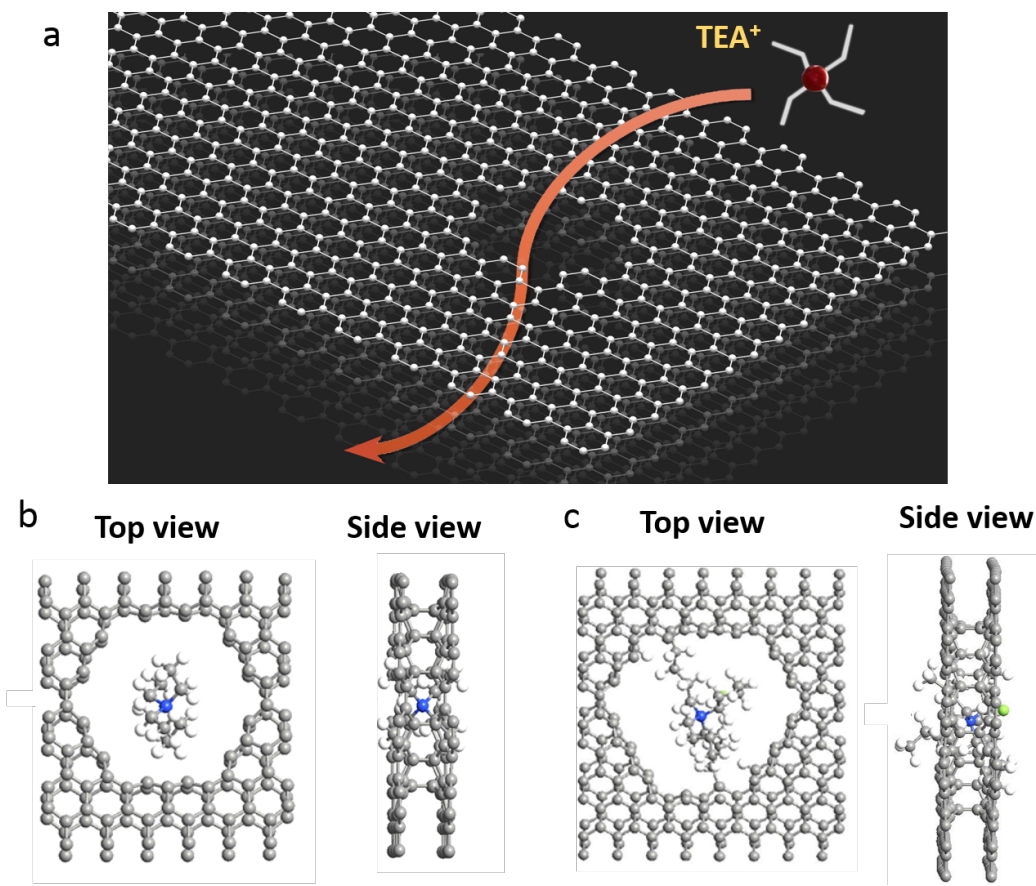

*(a) Defect-induced pores in FLG open otherwise inaccessible surface area by transporting electrolyte ions (e.g., tetraethylammonium ($TEA^+$)) to inter-layer gallery space. Density functional theory calculations showed that the intercalation of $TEA^+$ is more favorable (b) compared to tetra-n-butylammonium ($TBA^+$) (c).*

**Identification of best-suited electrolyte:**

Previously, it was observed that the best performance of SCs can be realized when the average micropore size in nanostructured bulk electrodes (e.g., carbide-derived carbon) matches the size of the ions in the electrolyte[8–15]. It is expected that such a resonant effect is true even for defect-induced pores in quasi-two dimensional FLG substrates (**Figure 1a**). Accordingly, in order to identify the best-suited electrolyte, we theoretically studied the effect of two different ions – tetraethylammonium ($TEA^+$) and tetrabutylammonium

(TBA$^+$), when they enter the defect-induced pores in FLG. The rationale in choosing these ions lies in the fact that organic electrolytes such as tetraethylammonium tetrafluoroborate (TEABF$_4$) and tetrabutylammonium hexafluorophosphate (TBAPF$_6$) exhibit a wider voltage range and yet are not highly expensive unlike ionic liquids. In our density functional theory (DFT) calculations, we started with an initial pre-optimized configuration of bilayer graphene with a 1 nm pore (See **Supplemental Figure S1**). Upon the introduction of TEA$^+$ ion into the interlayer spacing through the 1 nm pore, no significant changes were observed in the optimized bilayer geometry (**Figure 1b**) and the edge carbons did not show any chemical bonding/interactions with the TEA$^+$ ion. However, a large deformation in the structure of graphene sheets and an increase in the inter-layer spacing was observed in the presence of TBA$^+$ ion. In particular, we observed that the edge-carbons in the nanopore strongly interact with TBA$^+$ ion through chemical bonding preventing it from diffusing between graphene layers (**Figure 1c**). Such an observation may be rationalized in terms of the larger size of TBA$^+$ (~0.8 nm) compared to TEA$^+$ ion (~0.45 nm).[16]

**Figure 2: The experimental validation of DFT results.**

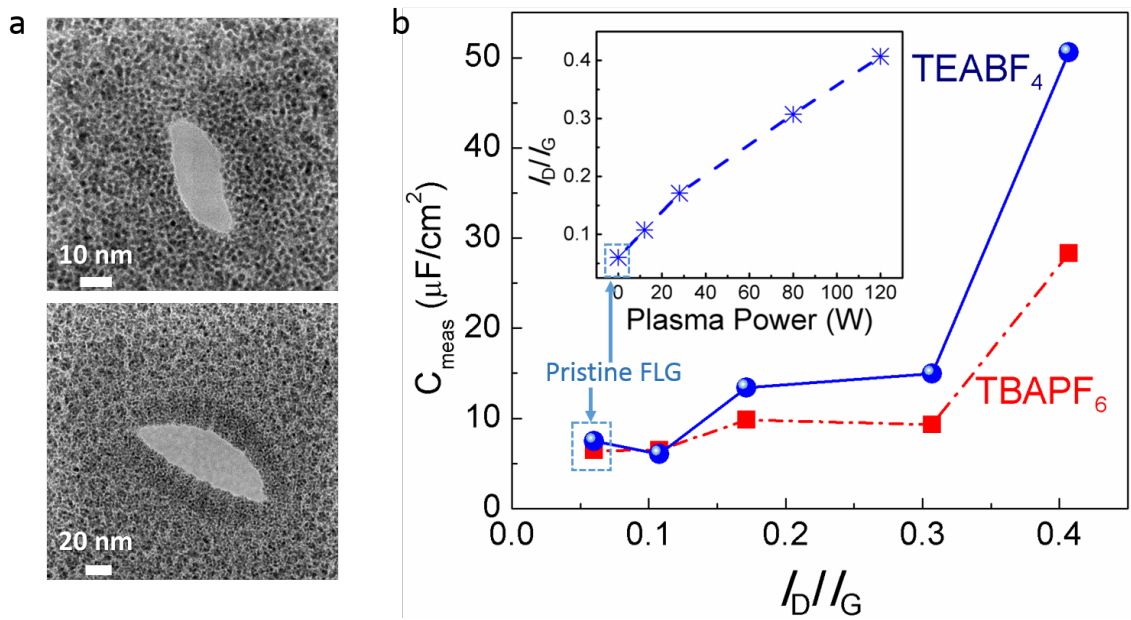

*(a) Transmission electron microscopy (TEM) images of the nanopores in FLG samples exposed to Ar$^+$ etched with a power of 120 W for 2 min. (b) The change in total measured capacitance ($C_{meas}$) as a function of defect densities (measured by $I_D/I_G$ ratio) for FLG*

*samples in the presence of: i) 0.25 M tetraethylammonium tetrafluoroborate (TEABF$_4$) in acetonitrile (blue dots and solid line), ii) tetrabutylammonium hexafluorophosphate (TBAPF$_6$) in acetonitrile (red squares and dash line). Inset: The ratio of intensity of D-band to the intensity of G-band ($I_D/I_G$) as a function of the plasma etching power shows a linear dependence.*

**Experimental validation of ion-pore size resonance effects**

We synthesized FLG layers and foams on Ni foil substrates through chemical vapor deposition (CVD) using previously described synthesis procedures (see **Methods**). We used Ar$^+$ plasma etching and N-doping to induce defects in FLG structures.[5] The defect-formation energy for extended defects is much lower compared to single- and di-vacancies, and thus we observed the formation of nanosized pores in FLG upon Ar$^+$ exposure (**Figure 2a**). The defects in graphene can be quantified using the normalized intensity ratio ($I_D/I_G$) of the well-known disorder or *D*-band to the graphitic or *G*-band in its Raman spectrum. We varied the Ar$^+$ plasma power (see **Methods** section) to produce FLG with different defect densities or $I_D/I_G$. As it may be expected, $I_D/I_G$ was found to increase linearly with the Ar$^+$ plasma power (see inset of **Figure 2b**). We performed cyclic voltammetry (CV) using a three-electrode electrochemical cell to study the effects of different electrolytes viz., 0.25 M TEABF$_4$, and TBAPF$_6$ dissolved in acetonitrile (ACN). The presence of $C_Q$ in series with $C_{dl}$ is expected to result in a total measured capacitance $C_{meas} = (C_{dl}^{-1} + C_Q^{-1})^{-1}$. We found that $C_{meas}$ in defected FLG increased for both electrolytes, which may be rationalized in terms of the higher $C_Q$ (= $e^2$*DOS ($E_F$), where $e$ is 1.6 x 10$^{-19}$ C), resulting from defect-induced increase in DOS($E_F$).[5] Although $C_{meas}$ increased for both electrolytes, TEABF$_4$ showed a much higher enhancement than TBAPF$_6$ (almost twice for the highest defect concentration). Such an observation concurs with our DFT calculations, which showed that TEA$^+$ ions are more suitable for accessing interlayer spacing in FLG through nanopores.

**Figure 3: The influence of N-doping on the electronic density of states**

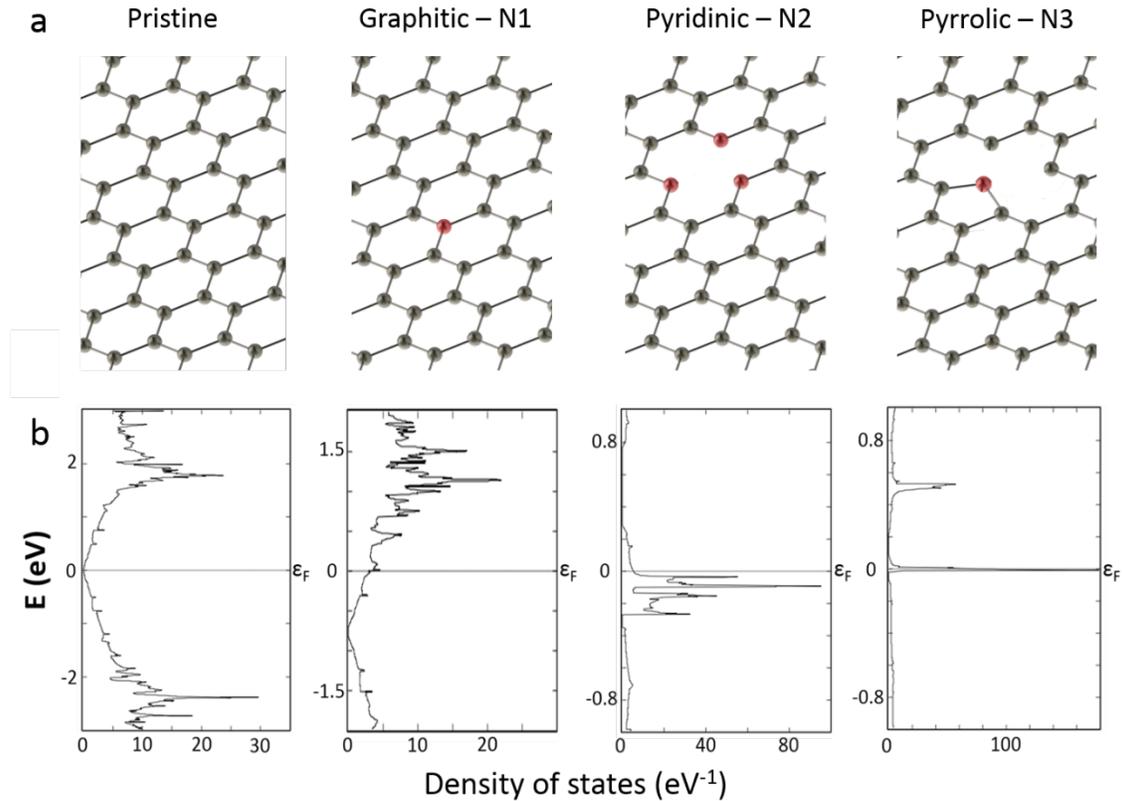

*(a) A schematic of different N-dopant configurations in graphene. (b) Density functional theory calculations show that the density of states (DOS) for pristine, graphitic, pyridinic, and pyrrolic samples (5x5 unit cells). The DOS at the Fermi level (0 eV) is negligible for pristine sample while it is very high for pyrrolic samples.*

**N-doping for improved power and energy density**

While the plasma etching of FLG significantly improves the measured capacitance and the energy density ($=0.5C_{meas}V^2$, where $V$ is the voltage), the increased $I_D/I_G$ compromises the power density due to increase in the resistance (see **Supplemental Figure S2**). Furthermore, the presence of pores in graphene severely weakens its structural integrity and thereby deteriorates the durability of the electrodes, as will be discussed later in **Figure 5**. Alternatively, the rich chemistry between carbon and nitrogen could be used to introduce N-dopants into graphene lattice in order to increase $C_Q$, $C_{meas}$, and energy density similar to nanopores, and yet retain intrinsic electrical conductivity and structural integrity of graphene. We prepared three different N-doped FLG structures using the CVD method (see **Methods**) for achieving SCs with high energy and power densities. As shown in **Figure 3a**, N-dopants can be found in at least three different configurations viz., graphitic, pyridinic, and pyrrolic. We analyzed the configuration stability, electron density, and DOS profiles (see **Figures 3b**) for all configurations using DFT calculations. Our previous work shows all the configurations exhibited positive formation energy values (i.e., energy released upon the formation of the structure from free atoms) suggesting that the doped sheets are stable.[17] The introduction of the dopants changed the symmetry of the lattice and resulted in drastically different DOS($E_F$) for all the three configurations (**Figure 3b**). Specifically, the DOS($E_F$) for pristine sample is negligible due to its semi-metallic nature while all other samples showed non-zero DOS($E_F$), with a very high DOS($E_F$) for pyrrolic configuration. The increase in DOS($E_F$) in addition to the fact that the pyrrolic type defects mimic the behavior of nanopores through extended defects (e.g., multiple vacancies) is useful for increasing both energy and power density of SC electrodes.

**Characterization of N-doped FLG structures**

As shown in **Supplemental Figure S3,** the C 1s line in the X-ray photoelectron spectroscopy (XPS) data of FLG exhibited a peak maximum at the binding energy of 284.45 eV. Upon N-doping, the C 1s line was observed to broaden and upshift (0.15–0.35 eV). The presence of different doping configurations was confirmed by deconvolution of the N 1s line using Voigtian components. The peak located at 401.5 eV (**Supplemental Figure S3b**) is identified with substitutionally doped nitrogen in the graphitic bonding configuration (sample N1). The peaks present at 398.8 (in sample N2) and 400.4 eV (in sample N3) arise from nitrogen bonded in the non-graphitic configuration, and were previously attributed to pyridinic and pyrrolic doping configurations respectively.[18] Based on our XPS studies, we estimated the N-dopant concentrations in our samples to be ~2 at.%. In addition to XPS confirmation, the Raman spectrum of N-doped graphene (**Supplemental Figure S4**) showed clear evidence for intense *D* & *D'*-bands for N2 and N3 samples, unlike the graphitic dopants in N1, due to the presence of extended defects and vacancies. Previously, we showed that the electron and phonon renormalization in

N-doped graphene increases the Fermi velocity ($v_F$) and thereby influences lattice vibrations locally near a dopant.[18] Indeed, a combination of micro-XPS and micro-Raman spectroscopy revealed that the local renormalization effects in N-doped graphene resulted in an effectively downshifted Raman 2*D* band (**Supplemental Figure S5**) with a large shift for N2 and N3 samples and a negligible shift of N1 in agreement with our previous studies.

**Figure 4: N doping leads to increased capacitance**

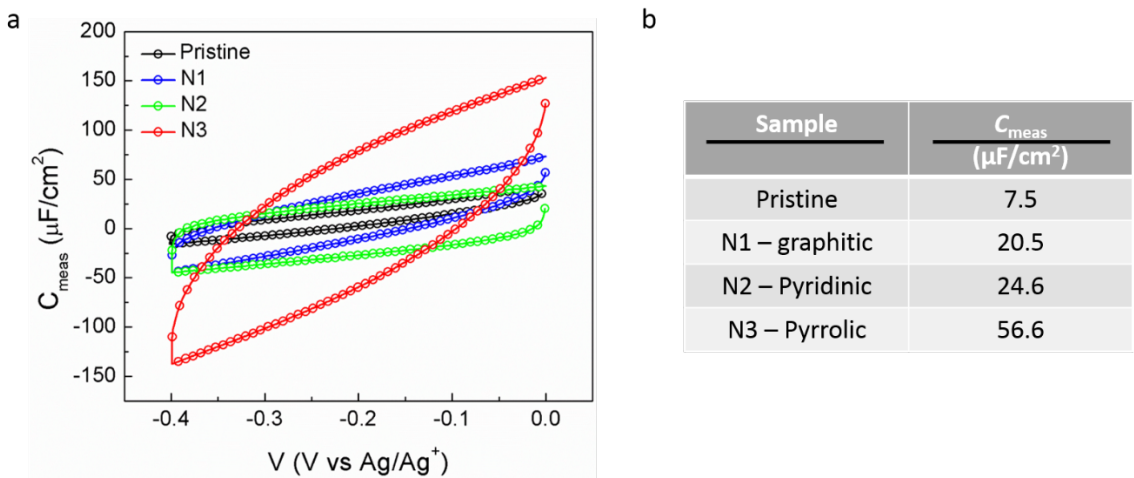

(a) Cyclic voltammetry (CV) curves (normalized by scan rate = 1000 mV/s) for pristine and different N-doped FLG samples obtained in 0.25 M tetraethyl ammonium tetrafluoroborate (TEABF$_4$) in acetonitrile. b) The total value of $C_{meas}$ for different FLG structures derived from (a).

**Electrochemical characterization of N-doped FLG**

For elucidating the influence of dopant configuration on $C_{meas}$, we performed CV measurements on the pristine, N1, N2, and N3 samples in the three-electrode setup with previously identified 0.25 M TEABF$_4$ in ACN. As shown in **Figure 4a**, the absence of redox peaks in the CV plots indicates the lack of specific reactions and pseudocapacitance arising from charge transfer at N-doped graphene /electrolyte interface. Clearly, we observed that the $C_{meas}$ values for N-doped FLG samples were higher than that of pristine FLG, and sample N3 (pyrrolic) was significantly different from samples N1 and N2, as predicted by our DFT calculations (see **Figure 4b**). We measured at least 5 different sets of samples to confirm the results presented in **Figure 4**. The controlled growth of a specific dopant configuration is highly difficult. Thus, to validate the hypothesis that the increase in $C_{meas}$ originates from pyrrolic configuration, we annealed the plasma etched FLG samples with nanopores in a 1" quartz tube furnace at 400 °C for 1 hr in Ar bubbled through ACN (see **Methods** for more details) into the furnace or only Ar. Interestingly, we found that the Ar-annealed FLG exhibited a slight decrease in $C_{meas}$ due to decrease $I_D/I_G$ upon annealing, as shown in **Supplemental Figure S6**. However, Ar-ACN annealed FLG structures showed a marked increase in $C_{meas}$ and concomitant decrease in resistivity due to the introduction of N-dopants. It is notable that pristine FLG structures annealed in Ar or Ar-ACN did not show any changes in $C_{meas}$ suggesting that the initial defects in the form of nanopores facilitated the introduction of N-dopants. Our detailed XPS characterization of Ar-ACN samples also revealed the presence of non-graphitic dopants similar to CVD grown samples (see **Supplemental Figure S7**).

**Figure 5: N-doped graphene foam-based coin cells with high-energy and power-densities.**

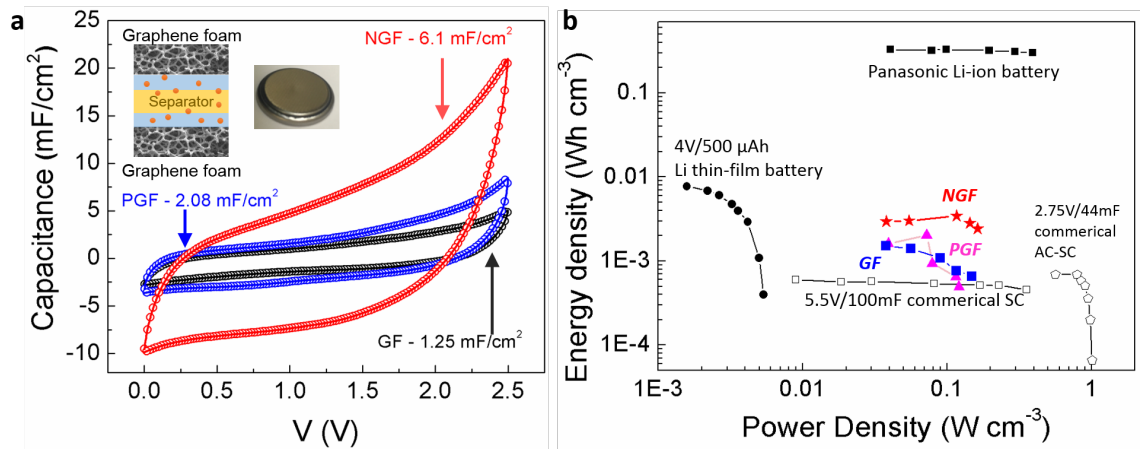

*(a) Cyclic voltammetry (CV) curves (normalized by scan rate = 1000 mV/s) for pristine, PGF and NGF coin cell devices obtained in 0.25 M tetraethyl ammonium tetrafluoroborate (TEABF4) in acetonitrile (ACN). (b) Ragone plot comparing the performance of GF coin cell devices with 0.25 M TEABF$_4$-ACN electrolytes to*

*conventional supercapacitors, Li-thin film batteries, and other energy storage devices.*[19,20]

**Realization of SC devices with high energy and power densities**

Building on the improved performance of the N-doped FLGs, we successfully constructed a coin-cell made of CVD grown graphene foams. It should be noted that SC devices based on CVD grown graphene have been limited to substrate-based micro-capacitors.[19,21] For the first time, to overcome this limit for bulk device applications, we synthesized large quantities of pristine FLG foams on Ni foam using CVD method (see **Methods**). The obtained pristine graphene foams (GFs) were etched through the Ar$^+$ plasma processing (for convenience, referred to as PGFs in **Figure 5**) at 120 W for 2 min for inducing nanopores (*cf.* **Figure 2**) and subsequently subjected to a post doping process through Ar-ACN annealing for producing N-doped GFs (NGFs in **Figure 5**). We fabricated symmetric coin cells with GF electrodes with a polymer separator placed in between the two electrodes. The GF electrode-separator-electrode sandwiched structures were assembled in a coin cell apparatus, as shown in the inset of **Figure 5a**. The CV profiles of the as-assembled coin cells (with three different electrodes viz., pristine GFs, PGFs, and NGFs) with 0.25 M TEABF$_4$ electrolyte are shown in **Figure 5a**. Clearly, while the PGFs showed a two-fold increase in $C_{meas}$ over pristine GFs, NGFs exhibited a six-fold increase. Galvanostatic charge-discharge measurements were carried out to calculate the energy and power density of the coin cells using $W = 1/2 C_{meas} V^2$, where $W$ is the energy density and $V$ is the cell voltage, and $P = dW/dt$, where $P$ is the power density and $dt$ is the discharge time. The charge–discharge characteristics of N-doped GF coin cells at 5 mA/cm$^2$ are shown in **Supplemental Figure S8**. The specific energy and power densities are calculated by normalizing the $W$ and $P$ value by the total coin cell volume of the two electrodes shown in Ragone plot (**Figure 5b**). The energy and power density values (**Figure 5b**) for the pristine and PGF devices are on par with the traditional SCs (e.g., commercially available activated carbon SCs), indicating our CVD grown GF is applicable in real energy storage devices. As shown in the Ragone plot in **Figure 5b**, the power density of PGF cells did not show much improvement and is indeed lower than pristine GF cells at low energy density due to increased resistance. On the other hand,

NGF devices exhibited a great enhancement of energy density (~5 times over activated carbon SCs) without compromising power density, which brings them closer to Li-ion thin film batteries in performance.[19] We observed that the performance of the SC coin cells comprised of NGFs was stable for over ten thousand cycles (unlike PGFs) with a drop of ~35% (**Supplemental Figure S8**).

**Discussion**

Defects are often perceived as material performance limiters. Contrary to this established notion, we demonstrated that the appropriate defect configuration could indeed alleviate roadblocks in harnessing the true energy storage potential of FLG. From a theoretical standpoint, defects in FLG break the crystal symmetry and thereby change the DOS($E_F$), which in turn significantly increases $C_Q$. The increase in $C_Q$ is an intrinsic limitation that impedes FLG-based SCs from delivering high-energy density. In addition to the increase in $C_Q$, the choice of the right electrolyte could enable the transport of ionic species through nanopores allowing access to otherwise unused interlayer spaces in FLG. In this work, we used a simple plasma etching process to induce defects in CVD-grown FLG structures and experimentally demonstrated an increase in $C_{meas}$. Although the $C_{meas}$ in defected FLG is increased by 250% (from 21 µF/cm$^2$ to ~50 µF/cm$^2$), the presence of defects in FLG severely weakens the structural integrity and compromises the power density. Indeed, we found that the defected FLGs cannot be cycled beyond 100-500 cycles, which is a serious limitation for SC devices. To overcome this challenge, we used the rich carbon and nitrogen chemistry to induce N-dopants into graphene lattice. Our comprehensive characterization and theoretical calculations showed that the non-graphitic dopants increase $C_Q$ by changing DOS($E_F$) and yet retain the necessary electrical conductivity and structural integrity. In case of pyrrolic N-dopants, we achieved a significant enhancement in $C_{meas}$ ~56 µF/cm$^2$ without using any pseudo-capacitive materials. The validity of our results is further reinforced by the experiments on ACN annealed defected FLG structures. The nanopores in defected FLG acted as a site for incorporating N-dopants into the graphene lattice and thereby exhibited an increase in $C_{meas}$, conductivity, and durability similar to CVD grown in situ doped N-doped FLGs.

While these results are exciting from a fundamental physics perspective, these developments would be futile if they cannot be extended to real-time devices with clear scalable manufacturing strategies. In this regard, we synthesized GFs using CVD in bulk quantities to produce coin cell devices. We induced pores in GFs through plasma etching and subsequently annealed these structures in ACN to achieve NGF-based coin cells with significantly higher energy and power densities than commercial SC devices. The CVD growth, plasma etching, and subsequent annealing are amenable for roll-to-roll production and are already being used for graphene production at industrial scales. Our results show new ways to tackle the inherent limitations of energy storage in nanocarbons by increasing $C_Q$ and accessible surface area through defect engineering without compromising the intrinsic properties of graphene, which opens a new paradigm for energy storage.

## Methods

### Computational details

The geometry optimization of bilayer graphene with electrolyte molecule was performed using DFT based first principle approach. The optimization was performed at mesh cut off 150 Rydberg. Local Density Approximation (LDA with Perdew-Zunger) has been used as exchange correlation functional with Double Zeta Polarized basis set. The structure was optimized until the force on every atom became less than 0.1eV/ Å. Sampling of Brillion zone for structure relaxation was taken as 1x3x3 using Monkhorst-Pack Scheme.

### Preparation of pristine and N-doped graphene

FLG layers and foams on Ni foil substrates were synthesized through chemical vapor deposition (CVD) using previously described synthesis procedures.[22] Atmospheric-pressure chemical vapor deposition (CVD) were used to synthesis and dope nitrogen atoms in pyridinic, pyrrolic, and graphitic configurations in FLG. Ni foils with thickness of 25 μm were placed away from the center of a 24 mm diameter tube furnace. The furnace was maintained at 900 °C under a flow of 200 sccm Ar and 120 sccm $H_2$. Ni foils were moved to center of the furnace after 90 minutes. Then the furnace was reset to 850°C and pristine graphene was synthesized by decomposing 10 sccm of methane for 10 min. Nitrogen doped graphene (see Fig. S1) was synthesized by flowing 50 sccm extra Ar through a mixture of benzylamine and acetonitrile (3:1, 1:1, and 0:1 for graphitic or N1, pyridinic or N2, and pyrrolic or N3 configurations respectively). Subsequently, methane flow was switched off and the samples were moved away from the center. The furnace temperature was cooled down to 400°C at 5°C/min. Then the $H_2$ flow was shut off and the furnace was maintained at 400°C for 90 min. Finally the samples were cooled to room temperature under Ar flow. For each configuration, at least three sister samples were used in our spectroscopic studies. FLG samples were also $Ar^+$ plasma etched using a reactive ion etching unit (Hummer 6.2) at multiple powers from 10-120 W for 2 min to induce defects. Dopant and defect concentration was quantified using Raman spectroscopy and X-ray photoelectron spectroscopy.

**Preparation of pristine and N-doped graphene foam**

The pristine graphene foam (GF) samples used in this study were grown on Ni foam substrate using thermal CVD technique similar to the FLG case. The nickel substrates were etched away by submersing the carbon/nickel foam in a 1:4 by volume Hydrochloric acid: nitric acid solution using concentrated acids. The recovered graphene foam was then rinsed with DI water and allowed to dry. Some samples were then subjected to a 120 W $Ar^+$ plasma for 2 minutes at 120 mtorr. Post nitrogen doping was accomplished by heating the foams in a tube furnace at 600°C for 60 minutes under a 500 sccm flow of argon which was bubbled through acetonitrile.

**Materials Characterizations**

A Dilor XY triple grating monochromator was used for collecting the micro-Raman spectra of all samples with the 532 and 633 nm excitation. X-ray photoelectron spectroscopy (XPS) studies were performed using a Kratos Axis Ultra DLD instrument and spectra were calibrated by C 1s at 284.6 eV. The morphology of the samples was observed using scanning electron microscopy (SEM, Hitachi S4800) with an accelerating voltage of 20 kV and tunneling electron microscopy (TEM, Hitachi H9500).

**Electrochemical measurements**

The electrochemical properties of samples were characterized in a Gamry reference 3000 electrochemical system. The electrolytes were 0.25 M tetraethylammonium tetrafluoroborate ($TEABF_4$, >99%) or tetrabutylammonium hexafluorophosphate ($TBAPF_6$, >99%) in acetonitrile (ACN). Two systems were used for electrochemical characterization: 1) a 3-electrode setup for single electrode characterization and 2) a 2-electrode cell (coin cell apparatus, MTI Corp) for symmetric supercapacitor measurements. In the 3-electrode cell, the FLG on Ni foil substrates were used as working electrodes, a Pt mesh was used as the counter electrode and a silver/silver ion electrode ($Ag/Ag^+$) was used as the reference electrode and. In the 2-electrode cell, EDLC devices were tested using symmetric GF samples as the electrodes with a Celgard (2325) trilayer separator. Each electrode and separator were soaked overnight (~20 hours) in 0.25 M TEABF$_4$-ACN electrolyte prior to the cell assembly. Cyclic voltammetry (CV) were measured from -0.4 to 0 V (0 – 2.5 V) for FLG (GF) samples with scan rate of 1000 mV/s. The electrochemical impedance spectroscopy (EIS) measurements were carried out with a perturbation signal of 10 mV in frequency range of 100 kHz to 0.1 Hz.

**Acknowledgements:** R.P. is thankful to Clemson University for providing start-up funds. A. M. R thanks the support from the National Science Foundation CMMI-Scalable Nanomanufacturing award. The authors thank Haijun Qian and Achyut Raghavendra for their help with transmission electron microscopy and high-resolution schematics.

**Author Contributions:** R.P. and A. M. R designed and initiated the studies. J. Z., M.K., A.S.C performed synthesis and electrochemical characterization. Y. L. conducted XPS characterization. A. S. performed density functional theory studies. All the authors contributed to data analysis and paper writing.

**REFERENCES**

1.  Liu, C., Yu, Z., Neff, D., Zhamu, A. & Jang, B. Z. Graphene-Based Supercapacitor with an Ultrahigh Energy Density. *Nano Lett.* **10,** 4863–4868 (2010).

2.  Wang, Y., Shi, Z., Huang, Y., Ma, Y., Wang, C., Chen, M. & Chen, Y. Supercapacitor Devices Based on Graphene Materials. 13103–13107 (2009). doi:10.1021/jp902214f

3.  Zhu, Y., Murali, S., Stoller, M. D., Ganesh, K. J., Cai, W., Ferreira, P. J., Pirkle, A., Wallace, R. M., Cychosz, K. A., Thommes, M., Su, D., Stach, E. A. & Ruoff, R. S. Carbon-Based Supercapacitors Produced by Activation of Graphene. *Science (80-. )*. **332,** 1537–1541 (2011).

4.  Stoller, M. D., Park, S., Zhu, Y., An, J. & Ruoff, R. S. Graphene-based ultracapacitors. *Nano Lett.* **8,** 3498–3502 (2008).

5.  Narayanan, R., Yamada, H., Karakaya, M., Podila, R., Rao, A. M. & Bandaru, P. R. Modulation of the Electrostatic and Quantum Capacitances of Few Layered Graphenes through Plasma Processing. *Nano Lett.* **15,** 3067–3072 (2015).

6.  Yamada, H. & Bandaru, P. R. Limits to the magnitude of capacitance in carbon nanotube array electrode based electrochemical capacitors. *Appl. Phys. Lett.* **102,** 1–5 (2013).

7.  Zhang, L. L., Zhao, X., Ji, H., Stoller, M. D., Lai, L., Murali, S., Mcdonnell, S., Cleveger, B., Wallace, R. M. & Ruoff, R. S. Nitrogen doping of graphene and its effect on quantum capacitance, and a new insight on the enhanced capacitance of N-doped carbon. *Energy Environ. Sci.* **5,** 9618 (2012).

8.  Shim, Y. & Kim, H. J. Nanoporous Carbon Supercapacitors in an Ionic Liquid: A

Computer Simulation Study. *ACS Nano* **4,** 2345–2355 (2010).

9. Simon, P. & Gogotsi, Y. Materials for electrochemical capacitors. *Nat. Mater.* **7,** 845–854 (2008).

10. Wu, P., Huang, J., Meunier, V., Sumpter, B. G. & Qiao, R. Complex Capacitance Scaling in Ionic Liquids-Filled Nanopores. *ACS Nano* **5,** 9044–9051 (2011).

11. Xing, L., Vatamanu, J., Borodin, O. & Bedrov, D. On the atomistic nature of capacitance enhancement generated by ionic liquid electrolyte confined in subnanometer pores. *J. Phys. Chem. Lett.* **4,** 132–140 (2013).

12. Merlet, C., Péan, C., Rotenberg, B., Madden, P. a, Daffos, B., Taberna, P.-L., Simon, P. & Salanne, M. Highly confined ions store charge more efficiently in supercapacitors. *Nat. Commun.* **4,** 2701 (2013).

13. Feng, G. & Cummings, P. T. Supercapacitor capacitance exhibits oscillatory behavior as a function of nanopore size. *J. Phys. Chem. Lett.* **2,** 2859–2864 (2011).

14. Largeot, C., Portet, C., Chmiola, J., Taberna, P. L., Gogotsi, Y. & Simon, P. Relation between the ion size and pore size for an electric double-layer capacitor. *J. Am. Chem. Soc.* **130,** 2730–2731 (2008).

15. Chmiola, J., Yushin, G., Gogotsi, Y., Portet, C., Simon, P. & Taberna, P. L. Anomalous Increase in Carbon Capacitance at Pore Sizes Less Than 1 Nanometer. *Science (80-. ).* **313,** 1760–1763 (2006).

16. Arcila-Velez, M. R., Zhu, J., Childress, A., Karakaya, M., Podila, R., Rao, A. M. & Roberts, M. E. Roll-to-roll synthesis of vertically aligned carbon nanotube electrodes for electrical double layer capacitors. *Nano Energy* **8,** 9–16 (2014).

17. Anand, B., Karakaya, M., Prakash, G., Sankara Sai, S. S., Philip, R., Ayala, P.,

Srivastava, A., Sood, A. K., Rao, A. M. & Podila, R. Dopant-configuration controlled carrier scattering in graphene. *RSC Adv.* **5,** 59556–59563 (2015).

18. Podila, R., Chacón-Torres, J., Spear, J. T., Pichler, T., Ayala, P. & Rao, a. M. Spectroscopic investigation of nitrogen doped graphene. *Appl. Phys. Lett.* **101,** 123108 (2012).

19. Yu, D., Goh, K., Wang, H., Wei, L., Jiang, W., Zhang, Q., Dai, L. & Chen, Y. Scalable synthesis of hierarchically structured carbon nanotube-graphene fibres for capacitive energy storage. *Nat Nano* **9,** 555–562 (2014).

20. Wu, Z., Parvez, K., Feng, X. & Müllen, K. Graphene-based in-plane micro-supercapacitors with high power and energy densities. *Nat Commun* **4,** (2013).

21. Yoo, J. J., Balakrishnan, K., Huang, J., Meunier, V., Sumpter, B. G., Srivastava, A., Conway, M., Reddy, A. L. M., Yu, J., Vajtai, R. & Ajayan, P. M. Ultrathin planar graphene supercapacitors. *Nano Lett.* **11,** 1423–7 (2011).

22. Radic, S., Geitner, N. K., Podila, R., Käkinen, A., Chen, P., Ke, P. C. & Ding, F. Competitive Binding of Natural Amphiphiles with Graphene Derivatives. *Sci. Rep.* **3,** 2273 (2013).

# Supplemental Information

# Defect-engineered graphene for bulk supercapacitors with high-energy and power densities

Jingyi Zhu[1], Anthony Childress[1], Mehmet Karakaya[1], Anurag Srivastava[2], Ye Lin[3], Apparao M. Rao[1,*], and Ramakrishna Podila[1,*]

[1] Department of Physics and Astronomy, Clemson Nanomaterials Center and COMSET, Clemson University, Clemson, SC 29634, USA

[2] ABV-Indian Institute of Information Technology and Management, Gwalior (M.P.) 474010, INDIA

[3] Department of Mechanical Engineering, University of South Carolina, Columbia, SC 29208, USA

*Corresponding author: rpodila@g.clemson.edu, arao@g.clemson.edu

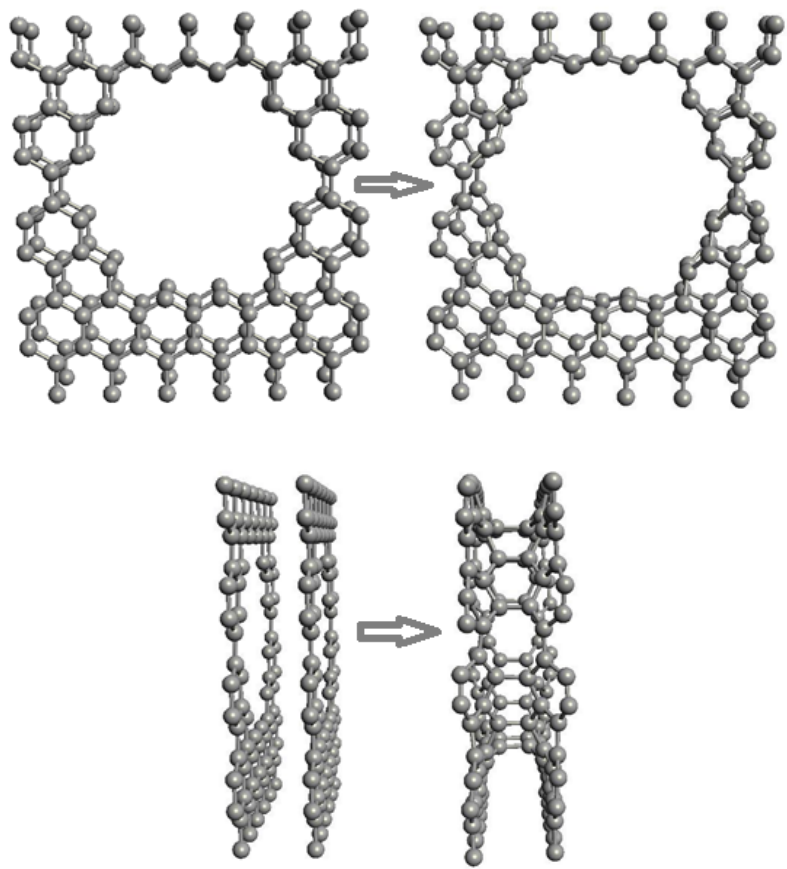

**Figure S1** Structure of bilayer graphene before and after optimization

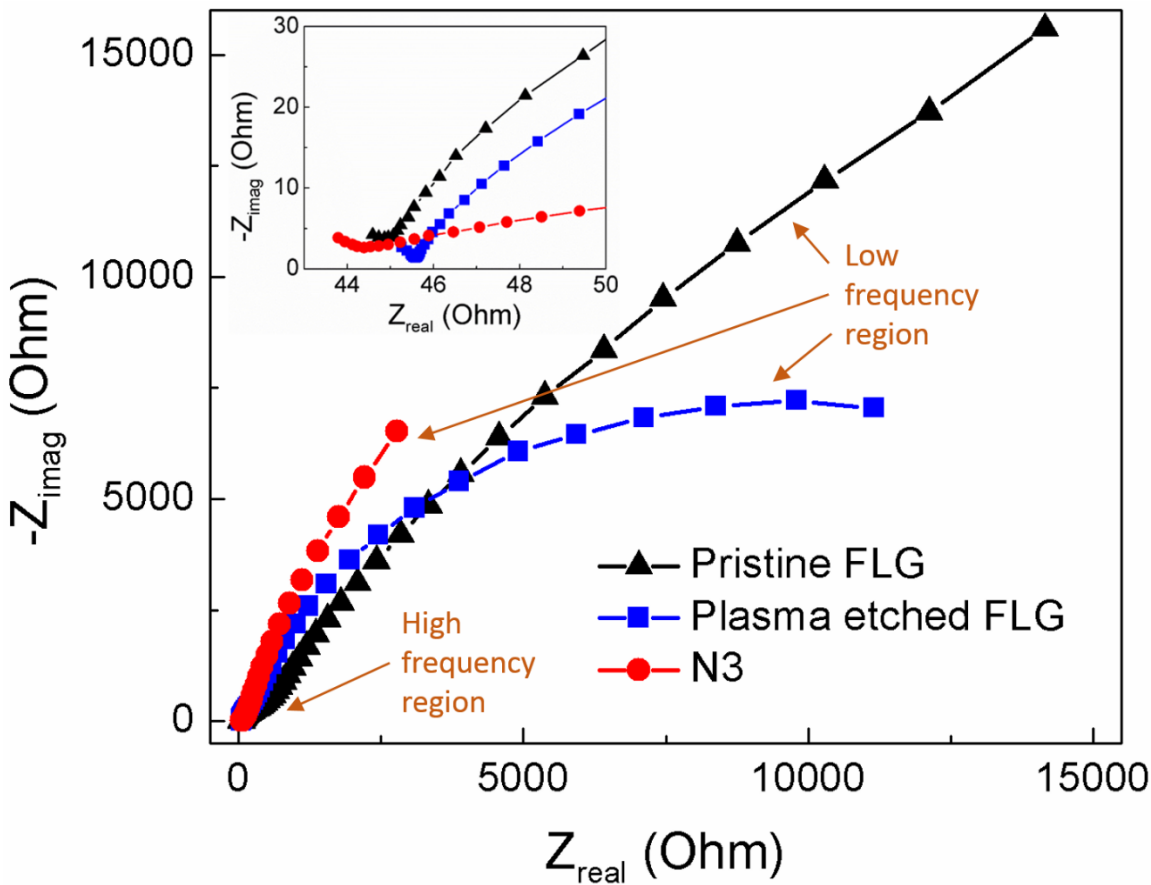

Figure S2 Nyquist plots of electrochemical impedance spectroscopy (EIS) of pristine FLG, Ar$^+$ plasma etched FLG, and N3 (or pyrrolic N-doped graphene samples) measured from 0.1 Hz to 100 kHz. Electrolyte: 0.25 M tetraethylammonium tetrafluoroborate (TEABF$_4$) in acetonitrile. Inset: Magnifies Nyquist plots for high frequency region. It can be seen that the plasma etched FLG has slightly higher equivalent series resistance (indicated by the first intercept of the Nyquist plots on the real axis[1]) and interfacial charge transfer resistance (presented by the radius of the semi-cycle at the high frequency region) from the high frequency region. Interestingly, the slope of data at the low frequency region, which depends on the electrolyte diffusion resistance (Warburg resistance $R_w$), is different for all three samples.[2,3] The higher slope indicates better ion diffusion within the electrodes.[4,5] Clearly, the plasma etched samples exhibit high

Warburg resistance, which could be attributed to the tortuous diffusion path of ionic species through defect-induced pores. However, sample N3 exhibits lower $R_w$ due to the presence of N-dopants in the pyrrolic configuration.

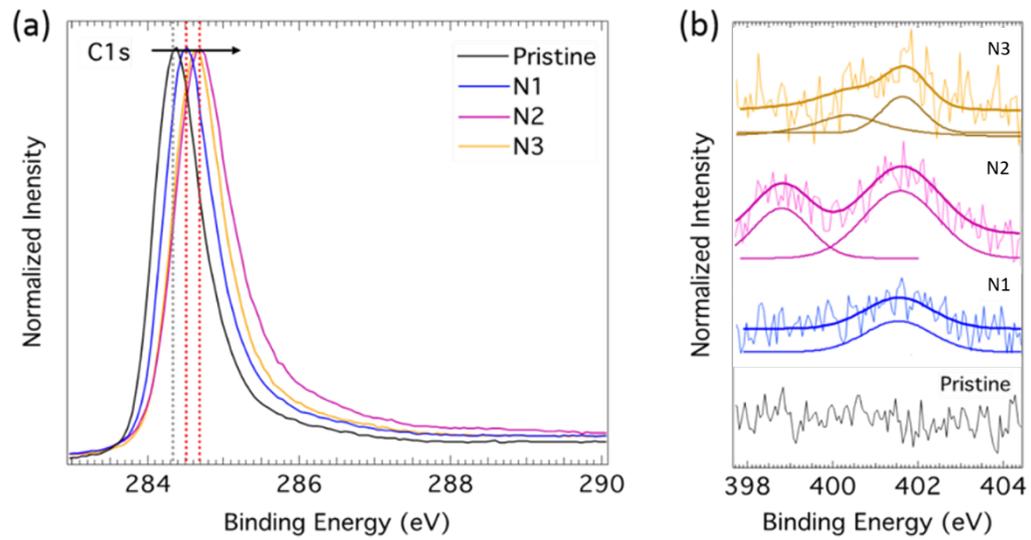

Figure S3  X-ray photoelectron spectroscopy for pristine and N-doped graphene, a) C 1s line, and b) N 1s line.

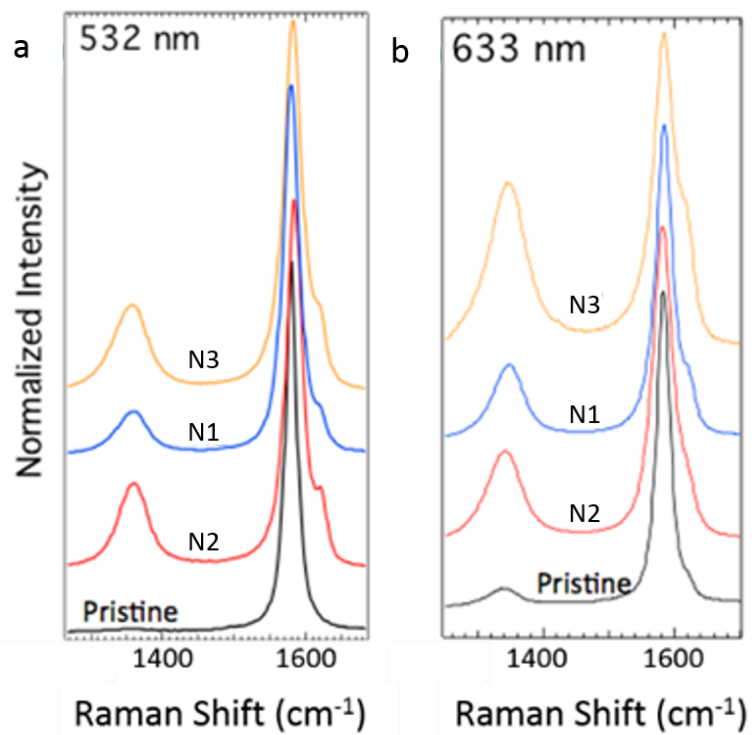

Figure S4   The *D* and *D'*-bands for pristine and N-doped few layer graphene grown using chemical vapor deposition, at excitations a) 532 nm and b) 633 nm. The *D* and *D'* bands are intense for samples N2 and N3 (non-graphitic doping configuration). The *D* band for N1 (graphitic doping configuration) is, however, less intense.

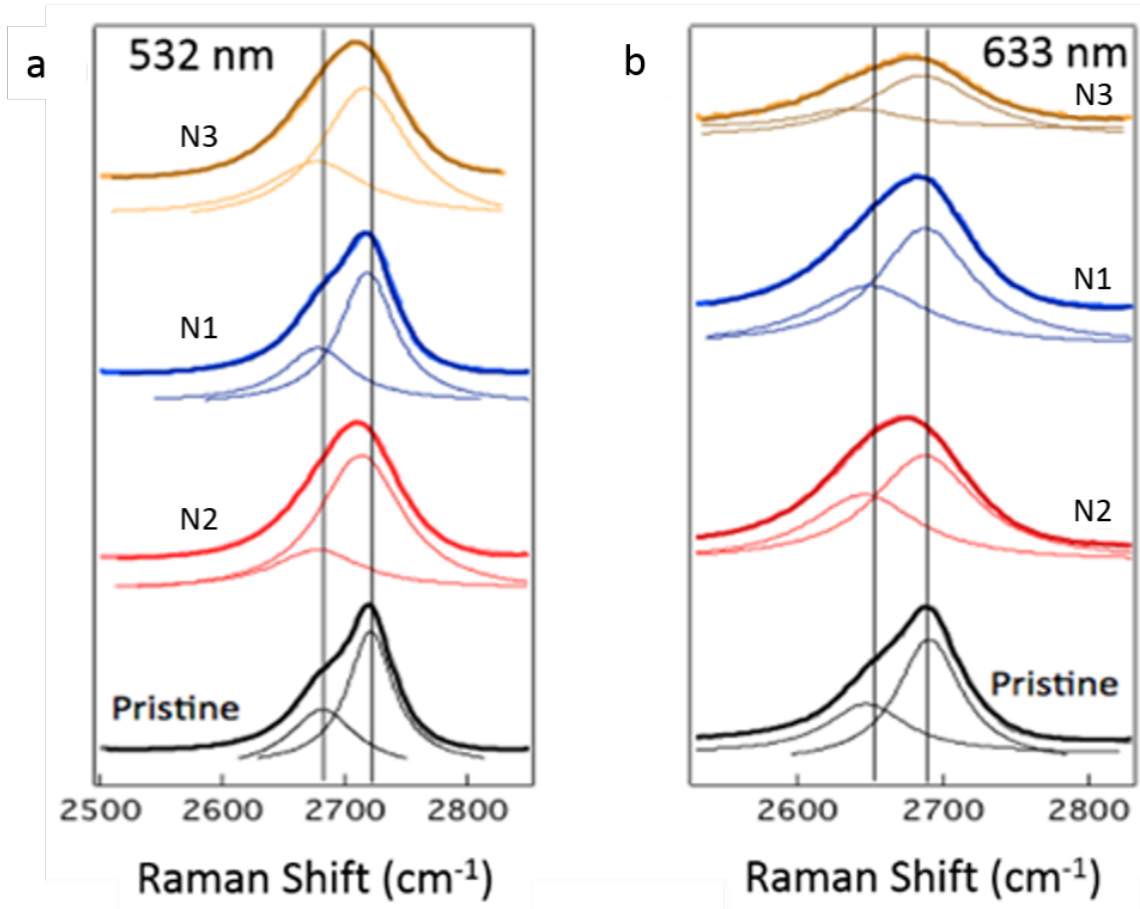

Figure S5    The 2D-band for pristine and N-doped few layer graphene grown using chemical vapor deposition, at excitations a) 532 nm and b) 633 nm. The electron-phonon-renormalization upon doping leads to a net down-shift in the 2D-band peak position for samples with a non-graphitic doping configuration. The traces below each spectrum indicate the deconvoluted peaks used for fitting. Clearly, sample N1 retains intense peaks in 2D-band with little downshift at both excitations.

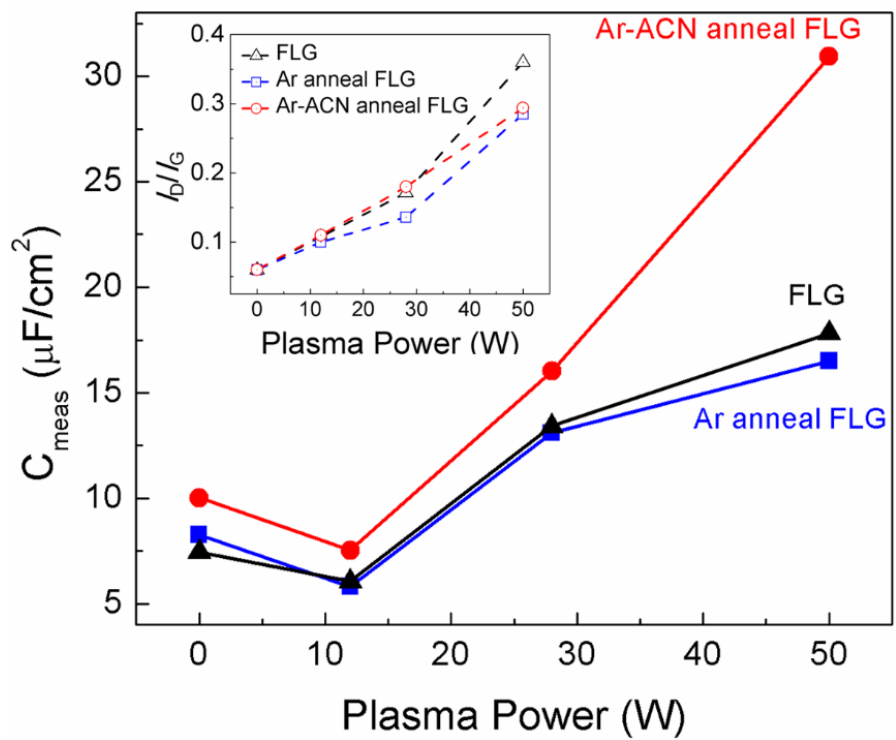

Figure S6    The change in total measured capacitance ($C_{meas}$) as a function of $Ar^+$ plasma power for FLG (black line and triangle dots), Ar annealed FLG (blue line and square dots) and Ar-ACN annealed FLG (red line and circle dots) samples in the presence of 0.25 M tetraethylammonium tetrafluoroborate (TEABF$_4$) in acetonitrile. Inset: The ratio of intensity of D-band to the intensity of G-band ($I_D/I_G$) as a function of the plasma etching power.

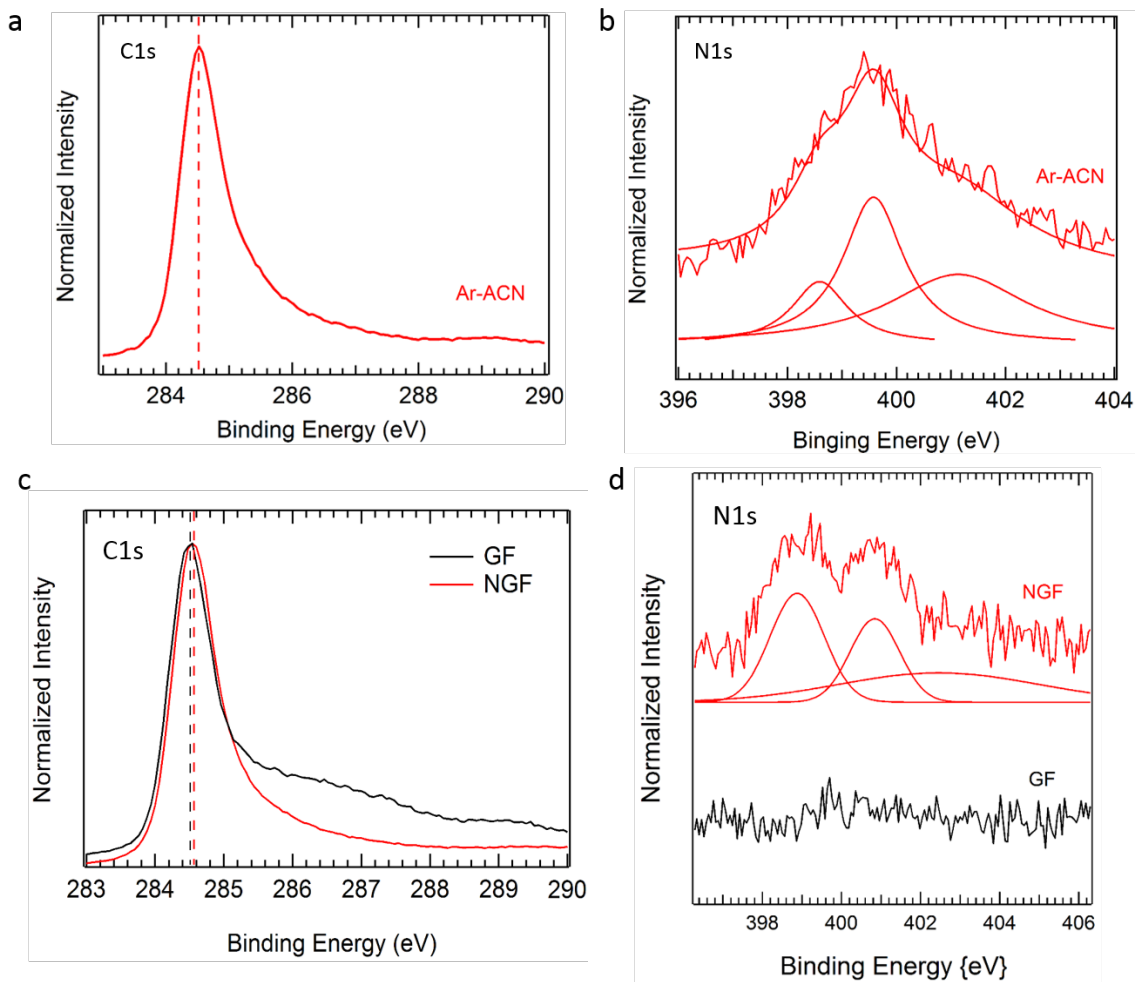

Figure S7  X-ray photoelectron spectroscopy for (a, b) Ar-ACN annealed FLG treated by 50 W Ar$^+$ plasma etching, and (c, d) pristine and N-doped graphene foams (GFs). We estimated the N-dopant concentrations in our samples to be ~2.5 at. %.

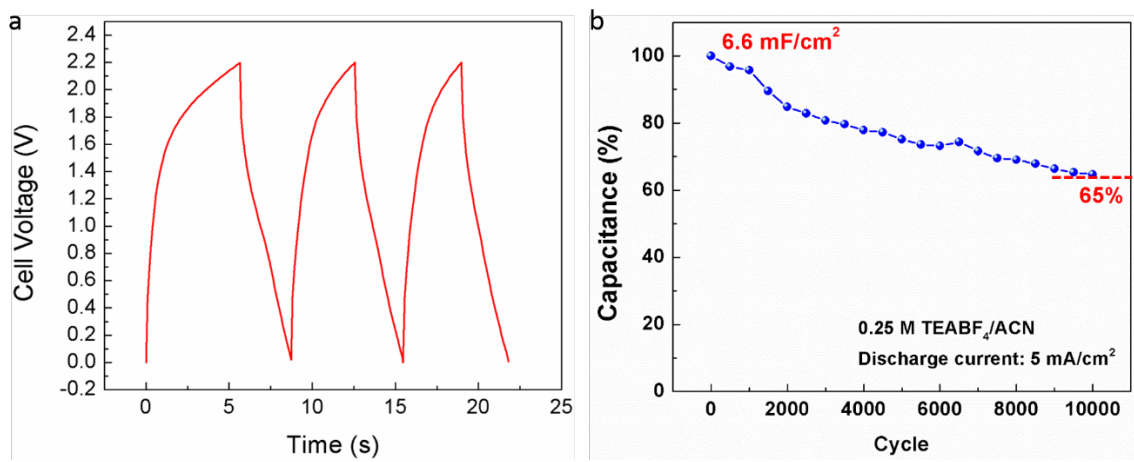

Figure S8  (a) The charge–discharge characteristics of N-doped GF coin cells at 5 mA/cm$^2$. (b)The coin cell device made from NGF electrodes exhibited good cycle stability with 35% degradation in performance over 10 000 cycles.

1.  Gamby, J., Taberna, P. L., Simon, P., Fauvarque, J. F. & Chesneau, M. Studies and characterisations of various activated carbons used for carbon/carbon supercapacitors. *J. Power Sources* **101,** 109–116 (2001).

2.  Jiang, Z., Jiang, Z., Tian, X. & Chen, W. Amine-functionalized holey graphene as a highly active metal-free catalyst for the oxygen reduction reaction. *J. Mater. Chem. A* **2,** 441–450 (2014).

3.  Sk, M. M. & Yue, C. Y. Layer-by-layer (LBL) assembly of graphene with p-phenylenediamine (PPD) spacer for high performance supercapacitor applications. *RSC Adv.* **4,** 19908–19915 (2014).

4. Stoller, M. D., Park, S., Zhu, Y., An, J. & Ruoff, R. S. Graphene-based ultracapacitors. *Nano Lett*. **8,** 3498–3502 (2008).

5. Zhang, K., Zhang, L. L., Zhao, X. S. & Wu, J. Graphene/Polyaniline Nanofiber Composites as Supercapacitor Electrodes. *Chem. Mater*. **22,** 1392–1401 (2010).